\documentclass[pre,showpacs,superscriptaddress,amsmathamsfonts,twocolumn,epsfig]{revtex4}
\usepackage{bm}
\usepackage{amsfonts}
\usepackage{times}
\usepackage[dvips]{graphicx}
\usepackage[T1]{fontenc}
\usepackage{color}
\usepackage{longtable}
\usepackage{epsfig}

\begin{document}
 \title{Anticipated synchronization in coupled complex Ginzburg-Landau systems}
\author{Marzena Ciszak}
\affiliation{CNR-Istituto Nazionale di Ottica, Florence, Italy}
\author{Catalina Mayol}
\affiliation{IFISC (Instituto de F{\'\i}sica Interdisciplinar y Sistemas Complejos), Campus UIB, Palma de Mallorca, Spain}
\author{Claudio R. Mirasso}
\affiliation{IFISC (Instituto de F{\'\i}sica Interdisciplinar y Sistemas Complejos), Campus UIB, Palma de Mallorca, Spain}
\author{Raul Toral}
\affiliation{IFISC (Instituto de F{\'\i}sica Interdisciplinar y Sistemas Complejos), Campus UIB, Palma de Mallorca, Spain}

\date{\today}

\begin{abstract}
We study anticipated synchronization in two complex Ginzburg-Landau systems coupled in a master-slave configuration. Master and slave systems are ruled by the same autonomous function, but the slave system receives the injection from the master and is subject to a negative delayed self-feedback loop. We give evidence that the magnitude of the largest anticipation time depends on the dynamical regime where the system operates (defect turbulence, phase turbulence or bichaos) and scales with the linear autocorrelation time of the system. Moreover, we find that the largest anticipation times are obtained for complex-valued coupling constants. We provide analytical conditions for the stability of the anticipated synchronization manifold that are in qualitative agreement with those obtained numerically. Finally, we report on the existence of anticipated synchronization in coupled two-dimensional complex Ginzburg-Landau systems.
\end{abstract}


\maketitle

\section{Introduction}
The synchronization of nonlinear dynamical systems is a topic of interest in many fields of science \cite{PRK01}. Particular attention has been payed to the synchronization of chaotic systems, both in unidirectional or bidirectional coupling configurations \cite{PCJ97,Boca02}. An interesting type of synchronization, so-called {\sl anticipated synchronization}, was proposed by Voss in \cite{voss1,voss2}. This author showed that, for particular parameter values, two identical chaotic systems unidirectionally coupled can synchronize in such a manner that the trajectories of the ``slave'' (the response system) anticipate (i.e. predict) those of the ``master'' (the sender system). Noticeably, the anticipation regime can be achieved without perturbing at all the dynamics of the master.

One of the coupling schemes proposed by Voss between the dynamics of the master, ${\bf x}(t)$, and slave, ${\bf y}(t)$, systems is given by the following set of equations:
\begin{eqnarray}
{{\bf{\dot{x}}} (t)} & = & {\bf F} ({\bf x} (t)),\label{eq1}\\
{{\bf{\dot{y}}} (t)} & = & {\bf F} ({\bf y} (t))+{\bf \cal K}\cdot \left[{\bf x} (t)-{\bf y}_{\tau}\right].\label{eq2}
\end{eqnarray}
Where $\bf{x}$ (``master'') and $\bf{y}$ (``slave'') are vectors of dynamical variables, $\bf{F}$ is a given vector function, $\tau$ is a delay time, ${\bf y}_{\tau} \equiv {\bf y}(t - \tau)$ is a delayed-feedback term in the dynamics of the slave, $\bf{\cal K}$ is a positive-definite matrix and the dot denotes a temporal derivative. For appropriate values of the delay time $\tau$ and strength of the elements of the coupling matrix $\bf{\cal K}$, it turns out that ${\bf y}(t) = {\bf x}(t+\tau)$ is a stable solution of Eqs.~(\ref{eq1}-\ref{eq2}). This condition can be interpreted as that the slave anticipates by a temporal amount $\tau$ the output of the master.

Since the seminal work by Voss, anticipated synchronization and its stability has been studied theoretically in several systems, including linear \cite{chialvo} and nonlinear \cite{chaotic} differential equations and maps \cite{maps}, as well as experimentally in e.g. semiconductor lasers with optical feedback \cite{refNewA} or electronic circuits \cite{refNewB}. The same phenomenon has been studied in excitable systems driven by noise \cite{marzena} where it was shown that the slave can predict the erratic generation of pulses originated by a random forcing in the master. In excitable systems, the existence of the anticipated solution has been related to the reduction of the excitability threshold induced by the coupling term in the slave system \cite{marzDyn,pyragas2010}. It was also shown that a sequence of many coupled systems can yield larger anticipation times \cite{r1}, although instabilities can appear if the number of coupled systems in the sequence is too large\cite{politi}. Theoretical studies have suggested that the mechanism of anticipated synchronization can play a role in a compensation of the conduction delays in coupled single neurons as well as in coupled excitable media \cite{zeroLag1,fernanda}. Such compensation may lead to the emergence of zero-lag synchronization between spatially separated brain regions, as observed in experiments \cite{singer1}. As a possible application, anticipated synchronization has led to the design of a \emph{predict-prevent} control method \cite{control1,control2} to avoid unwanted pulses in excitable or other chaotic systems. In this control method an auxiliary slave system is introduced to predict the firings of the master system, such that the information coming from the former is consequently transformed into a control signal that suppresses, if needed, those unwanted pulses of the master system.

So far, most examples and applications have considered systems with a small number of dynamical variables. It is the aim of this paper to go a step forward and show that anticipated synchronization can be achieved in spatiotemporal chaotic systems. To this end, we consider a master system described by the prototype complex Ginzburg-Landau equation to which we add a conveniently coupled slave system. We first consider the one-dimensional case and characterize numerically the parameter space for which the anticipated solution exists and is stable. By introducing a complex-valued coupling constant between master and slave, we find an increase of the anticipation time with respect to the case of a real-valued coupling constant \cite{pyragas2008}. Then, we show the existence of a relationship between the largest anticipation time and the linear autocorrelation time. We also consider a two-dimensional scenario and show numerically that anticipated synchronization can also be achieved in this case. Finally, we present the results of an approximate linear stability analysis that can reproduce some of the features observed in the numerical simulations.

\section{Model}
A well-known model equation which displays a rich variety of spatiotemporal dynamics is the complex Ginzburg-Landau (CGL) equation \cite{refNew1,refNew2}, which in one spatial dimension reads:
\begin{eqnarray}
\dot{A}&=&\epsilon A+ \alpha_1 A_{xx}- \alpha_2 |A|^2A
\label{eq3}
\end{eqnarray}
where $\alpha_1 = 1 + i c_1$ and $\alpha_2= 1 + i c_2$ are complex constants. Here $A = A(x,t)\equiv\rho(x,t)e^{i\phi(x,t)}$ is a complex field of amplitude $\rho$ and phase $\phi$, and $A_{xx}=\displaystyle\frac{\partial ^2 A}{\partial x^2}$ is the second-order derivative with respect to the space variable $0\leq x\leq L$, being $L$ the system length. $\epsilon $ is a control parameter inducing instability if it is positive, $c_2$ is a measure of the nonlinear dispersion and $c_1$ is the linear dispersion parameter. Equation (\ref{eq3}) admits plane-wave solutions of the form:
\begin{eqnarray}
A_q (x,t)&=&\sqrt{\epsilon-q^2}e^{i(qx+\Omega(q) t)}
\label{eq3a}
\end{eqnarray}
where $q$ is the wave number in Fourier space bounded by $-\sqrt{\epsilon}\leq q \leq \sqrt{\epsilon}$ and $\Omega(q)=
-c_2-(c_1-c_2)q^2$ is the dispersion relation. All plane-waves become unstable when crossing the so-called Benjamin-Feir or Newell line given by $c_1c_2=-1$ for $\epsilon=1$ \cite{shraiman}. Above this line different dynamical regimes were identified:
defect turbulence, phase turbulence, bichaos and spatiotemporal intermittency. Defect turbulence is a strongly disordered region
in which defects, as well as other localized structures, appear displaying a rich dynamics. Phase turbulence is a state
weakly disordered in amplitude and strongly disordered in phase whereas the bichaos region is an alternating mixture of phase and
defect turbulence states. In the spatiotemporal intermittency region stable traveling waves interrupted by turbulent bursts
exist. The complex Ginzburg-Landau equation is a universal model for the evolution of an order parameter describing the loss of stability of a homogeneous state through a Hopf bifurcation (and it is often called {\sl model A} in analogy to phase transitions). Thus, Eq. (\ref{eq3}) is the normal form for any system that is time-translational invariant and reflection symmetric in which a supercritical Hopf bifurcation appears. As an example, it can be derived from a model of bidirectionally coupled FitzHugh-Nagumo cells, where the membrane potential $v(t)$
is assumed to be $v(t)=A(t)\exp{(i\omega |A|^2t)}$
\cite{rabinovich}. In this case, at variance with the single cell, a chaotic behavior is possible
due to the additional degrees of freedom introduced by the
spatially extended excitable cells. In the particular case $c_1=c_2=0$, Eq.
(\ref{eq3}) reduces to the so-called real Ginzburg-Landau equation which describes superconductivity in the
absence of magnetic field. In the limit $c_1,c_2\rightarrow \infty$ the equation reduces to the nonlinear Schr\"{o}dinger equation with its well-known soliton solutions.

Following the general schemes, Eqs.(\ref{eq1}-\ref{eq2}), we study the situation in which two equations
are coupled in a master-slave configuration, such that the slave
system contains an input from the master system and a negative
self-feedback term. Namely, the system equations read:
\begin{eqnarray}
\dot{A}&=&\epsilon A+\alpha _1A_{xx}-\alpha _2|A|^2A\label{eq5}\\
\dot{B}&=&\epsilon B+\alpha _1B_{xx}-\alpha _2|B|^2B+\kappa (A-B_\tau)\label{eq6}
\end{eqnarray}
with a general complex-valued coupling constant $\kappa \equiv K e^{i\theta}$. Complex coupling terms have been previously considered in e.g. laser systems \cite{ikeda}. $A=A(x,t)$ is the master system, $B=B(x,t)$ the slave, and $B_\tau=B(x,t-\tau)$, being $\tau$ a constant delay time.  In what follows we set $\epsilon =1$. Our main results are presented in the next two sections. First, we describe in detail the numerical results for the one-dimensional master-slave configuration and then, to a lesser extent, examples of anticipated synchronization in the case of two-dimensional systems. Next, we develop a stability analysis that can roughly explain the numerical results for the one-dimensional case.

\section{Numerical results}
Since the anticipated synchronization regime $B(x,t)=A(x,t+\tau)$ is always an exact solution of Eqs.(\ref{eq5}-\ref{eq6}), the main point of interest is to determine its range of stability, i.e. the range of parameter values, in particular the maximum time delay $\tau$, for which this regime is reached asymptotically and independently of initial conditions. We expect that the stability of the anticipated synchronization solution would depend on the nature of the chaotic dynamics: the stronger the chaos, the smaller the anticipated region \cite{thomas}. Earlier work in
systems with a small number of degrees of freedom showed \cite{r1} that anticipated synchronization in chaotic systems exists for
those (small) delay times for which the first order linear approximation is valid to represent the delayed coupling scheme\cite{chialvo}. According to previous results \cite{r1}, we expect (and will show that this is indeed the case) that the largest anticipation time is related to the linear autocorrelation time estimated from the time series. Since the different dynamical regimes exhibited by the complex Ginzburg-Landau equations have different linear autocorrelation times, we expect the maximum anticipation time to decrease when moving from the (less chaotic) phase turbulence into the bichaos and (most chaotic) defect turbulence regimes.

\begin{figure}[h]
\includegraphics[width=0.7\linewidth]{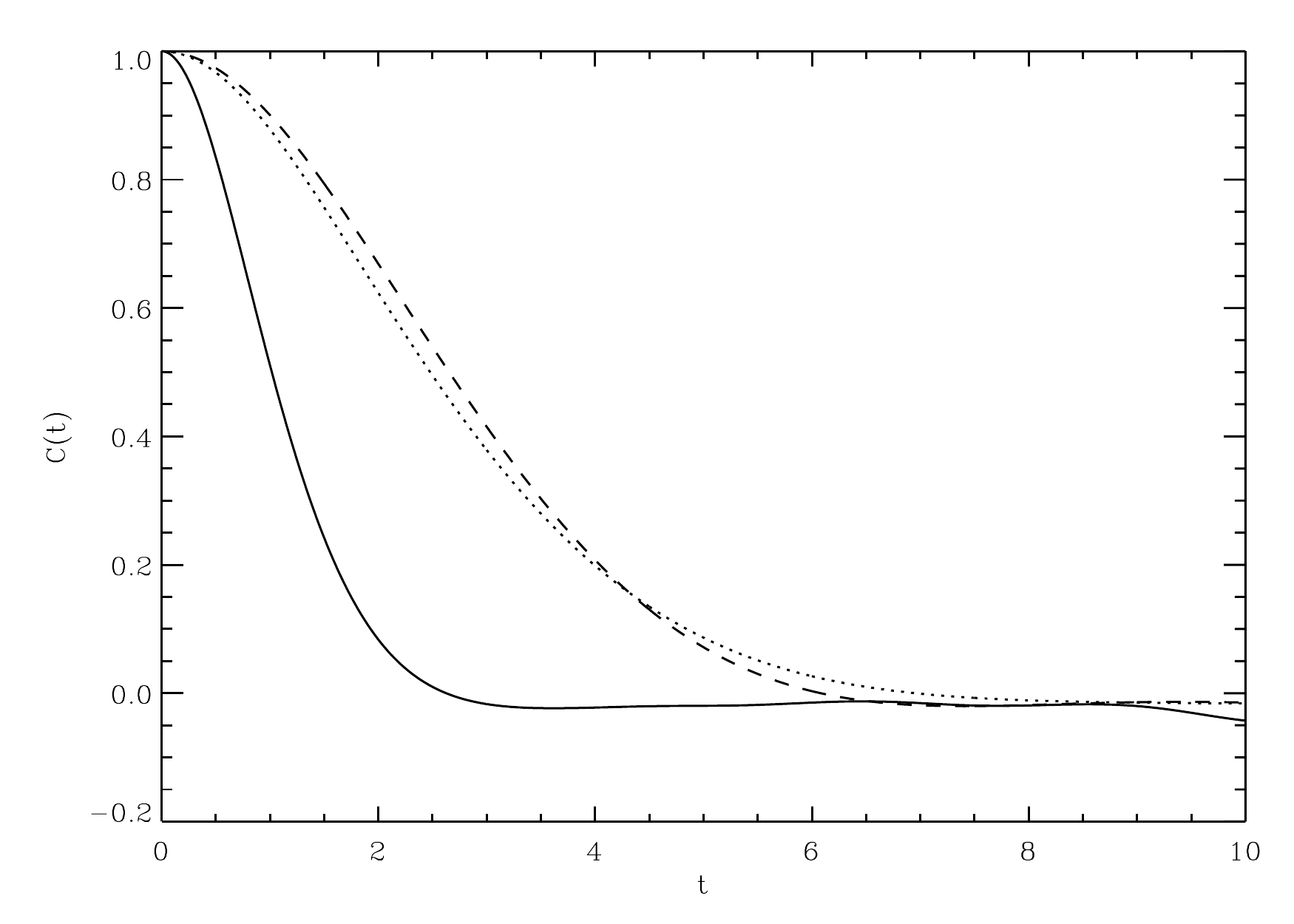}
\caption{\label{f1}
Autocorrelation function versus time for the three chaotic regimes
of the complex Ginzburg-Landau equation: defect turbulence (solid line), phase turbulence (dashed line) and bichaos (dotted line). The system parameters used are: $\epsilon =1$ (in all cases), $(c_1,c_2)=(3,-2.5)$ for defect turbulence, $(c_1,c_2)=(1.5,-0.9)$ for phase turbulence, and $(c_1,c_2)=(1.1,-1.2)$ for bi-chaos regime, as given in \cite{shraiman} }
\end{figure}

Let us start by analyzing the autocorrelation time of Eq.~(\ref{eq5}) in the different regimes explained in the previous section. In our numerical simulations \footnote{For the numerical integration of Eqs.(\ref{eq5}-\ref{eq6}) we have used a two-step method (``slaved leap frog" of Frisch \cite{frisch} with the corrective algorithm used in \cite{montagne1}) to integrate the Fourier modes, assuming periodic boundary conditions. The integration time step is $\delta t= 2 \times 10^{-3}$. We use random initial conditions, different in the master and the slave, in order to obtain independent initial dynamics in both systems. The size of the system is $L=N \delta x$, with $N=64$ and $\delta x=0.2$ in the defect turbulence regime, $\delta x=1.6$ in the bichaos regime and $\delta x=2$ in the phase turbulence regime.}, we computed the normalized autocorrelation function 
\begin{equation}
C(t)=\displaystyle\frac{\langle |A(x,s)A(x,t+s)|\rangle_\textrm{st}-\langle |A(x,s)|\rangle_\textrm{st}\langle |A(x,t+s)|\rangle_\textrm{st}}{\langle |A(x,s)^2|\rangle_{\textrm{st}}-\langle |A(x,s)|^2\rangle_\textrm{st}},
\end{equation}
where $\langle\dots\rangle_\textrm{st}$ denotes a time average over $s$ in the stationary state and, for the periodic boundary conditions considered here, the average is independent of the location of the point $x$. In Fig.\ref{f1} we plot the autocorrelation functions, evaluated at the particular point $x=L/2$, as a function of time for three cases corresponding to the defect turbulence, bichaos and phase turbulence regimes. The characteristic decay time of these functions can be quantified by the linear autocorrelation time $T_c$, computed from the numerical series as:
\begin{equation}
T_c\sim \frac{\delta t}{M}\sum_{i=0}^{M-1} \left[\ln \left(C(t_{i})/C(t_{i+1})\right)\right]^{-1},
\label{eqTc}
\end{equation}
where $t_i=t_0+i \delta t $ are the different points at which the
correlation function is computed. The upper limit of this sum
extends up to a time $t_M$ before the appearance of the small
oscillations at the tail of the correlation function, see Fig.
\ref{f1}.

Using this definition, we have obtained the following linear
autocorrelation times for the aforementioned parameter values:
$T_c=0.55$ (corresponding to the point in the defect turbulence
regime), $T_c=1.65$ (bichaos) and $T_c=1.76$ (phase turbulence).
From these values, and according to the discussion above, we
expect that the defect turbulence is the region in which
anticipated synchronization will be stable in a smallest region. Therefore, in the rest of the section, we concentrate in
the defect turbulence regime and determine the anticipated region
as a function of the time delay $\tau$ and the coupling parameters
$K$ and $\theta$.

\begin{figure}[h]
\includegraphics[width=0.7\linewidth]{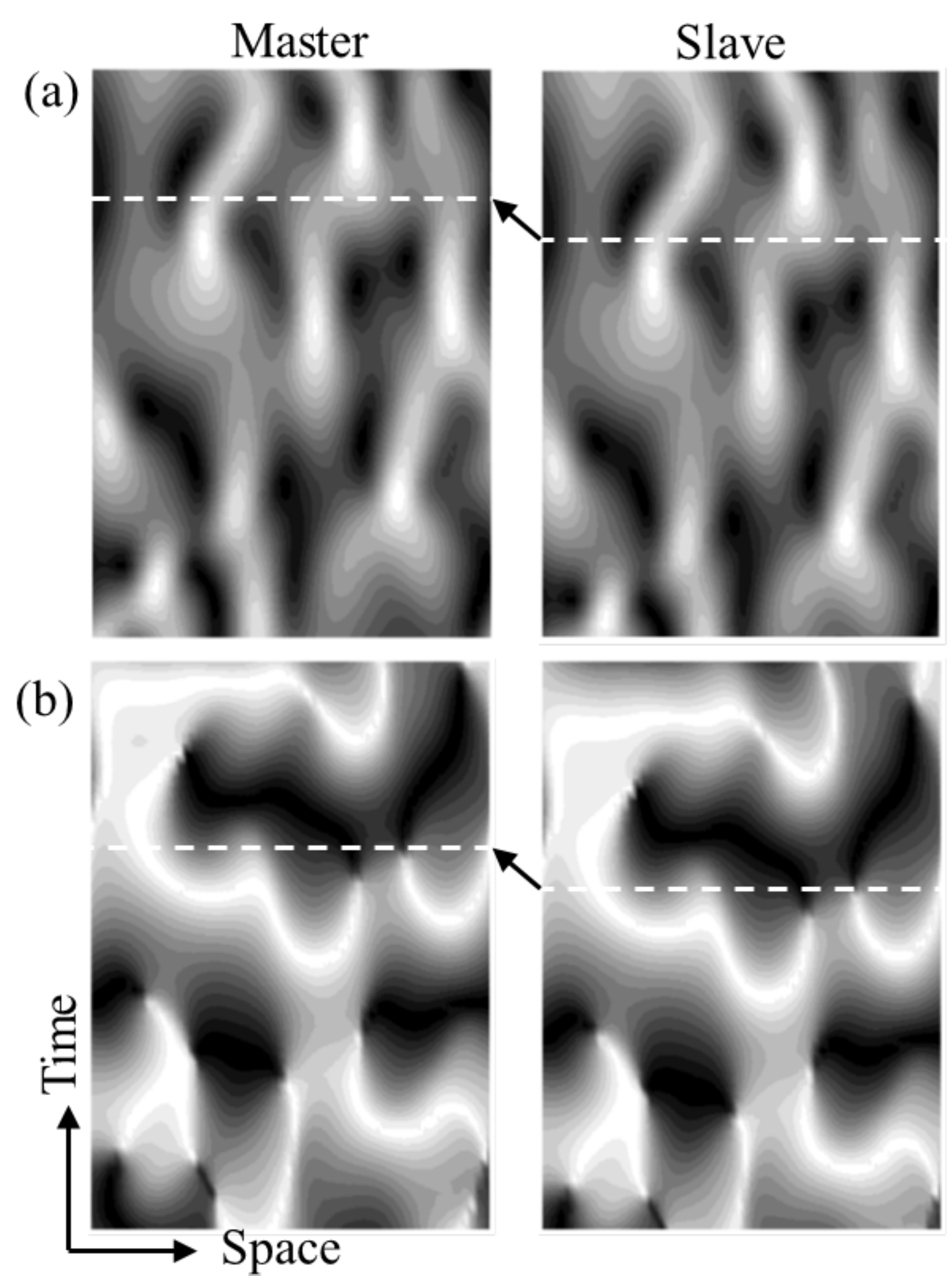}
\caption{\label{f2} Spatiotemporal dynamics of (a) amplitude and (b) phase of the master and the slave systems for two unidimensional complex Ginzburg-Landau systems coupled as in Eqs.(\ref{eq5}-\ref{eq6}). Parameter values: $\tau=0.6$, $\theta=\frac{\pi}{4}$, $K=0.6$, $N=64$, $\delta x=0.2$, $\delta t=2\times10^{-3}$, $c_1=3$, $c_2=-2.5$ (in a range of
defect turbulence regime). The difference in time between the horizontal dashed lines is the anticipation time $\tau=0.6$.}
\end{figure}

\begin{figure}[h]
\includegraphics[width=0.9\linewidth]{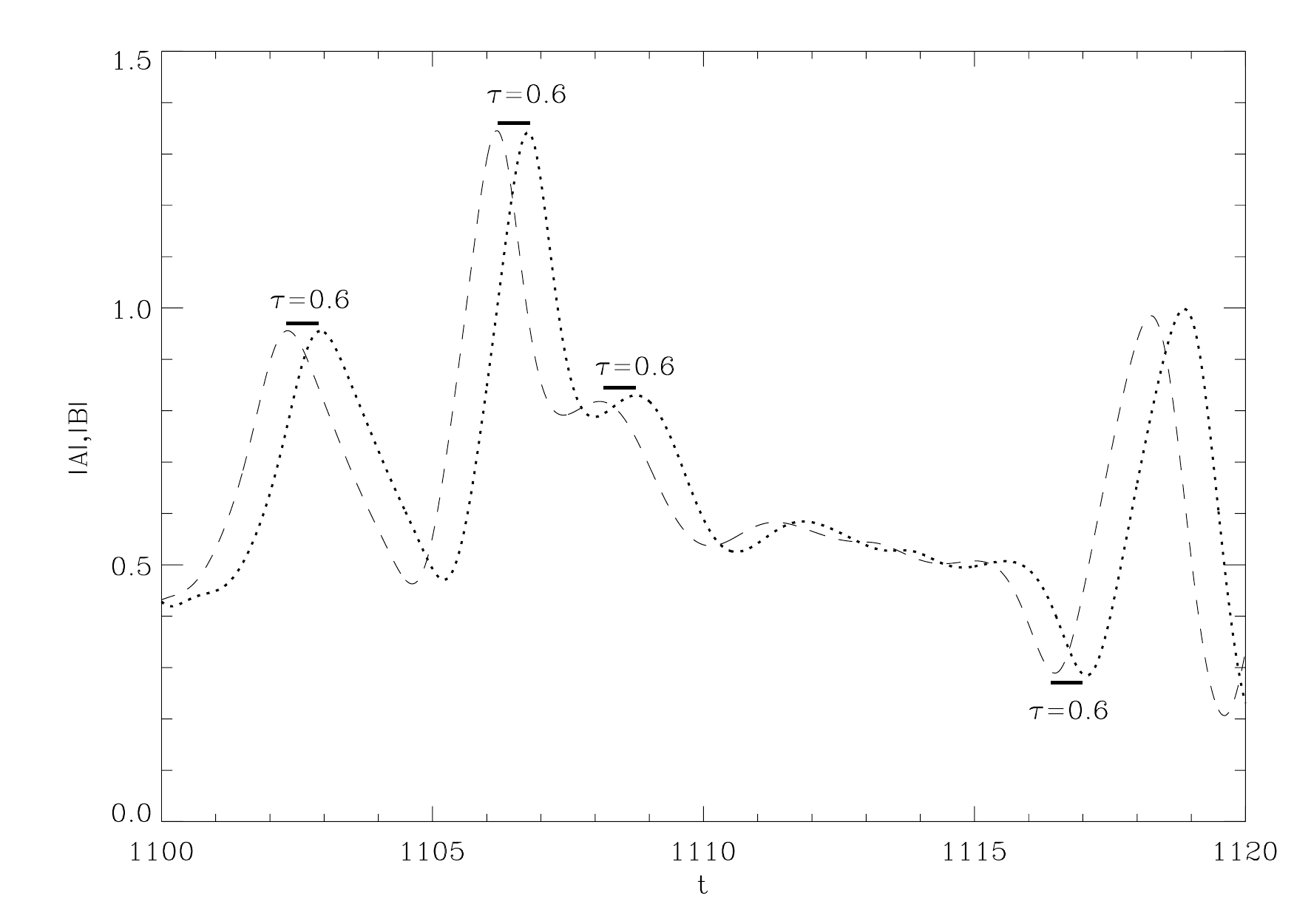}
\caption{\label{f3} Time series of the master amplitude $|A|$
(dotted line) and the slave $|B|$ (dashed line) for the complex Ginzburg-Landau equation in the
defect turbulence regime. Horizontal arrows mark the anticipation
time. Same parameters as in Fig.\ref{f2}.}
\end{figure}

We display in Fig. \ref{f2} spatiotemporal plots of the master and slave fields (amplitude and phase) in the defect turbulence regime, $(c_1,c_2)=(3,-2.5)$, for particular values of parameters $\tau=0.6$, $K=0.6$ and $\theta=\pi/4$ for which anticipated synchronization turns out to be stable. It can be clearly seen in this figure how, both in amplitude and phase, the field $A(x,t)$ of the master coincides with that of the slave at an earlier time $B(x,t-\tau)$. The same behavior can be seen in Fig.\ref{f3} where we plot the time evolution of the modulus of the master and slave fields in a particular point in space, $x=L/2$. As will be shown later, the value of $\tau=0.6$ is approximately the largest delay time for which anticipated synchronization is stable for the given values of $K$ and $\theta$. Indeed, as indicated in Fig.\ref{f3}, this maximum anticipation time approximately coincides with the linear autocorrelation time for these parameter values.

\begin{figure}[h]
\includegraphics[width=1.\linewidth]{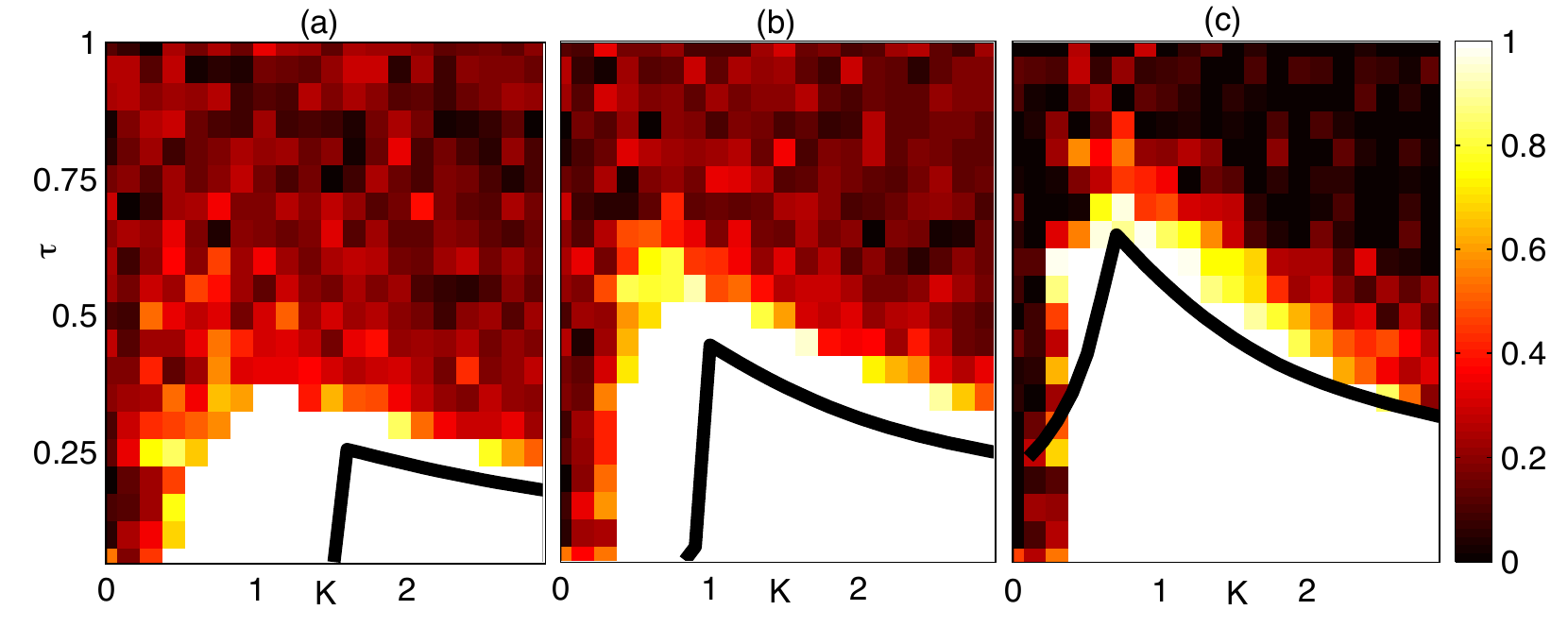}
\caption{\label{f4} (color online) Amplitude correlations between master $|A|$ and slave
$|B_\tau|$ time series in the defect turbulence regime with $(c_1,c_2)=(3,-2.5)$ at the spatial position $x=L/2$ for (a)
$\theta=-\frac{\pi}{6}$, (b) $\theta=0$ and (c)
$\theta=\frac{\pi}{6}$. White color corresponds to high correlations
($C>0.99$), meanwhile dark colors correspond to lower correlations
($C<0.99$). Solid black lines represent analytical results obtained from
Eq.(\ref{bifurcation}) using a linear stability analysis.}
\end{figure}

To determine the stable regions where anticipated synchronization can occur in the defect turbulence regime, we have performed extensive numerical simulations of Eqs.(\ref{eq5}-\ref{eq6}) scanning the $(\tau,K,\theta)$ parameter space. Some results are presented in Fig.\ref{f4}, where we plot, using a color scale, the normalized correlation coefficient between the master at time $t$, $A(x,t)$, and the slave at a time $\tau$ earlier, $B(x,t-\tau)$, for three different values of the coupling phase $\theta$. As it has been found in previous studies, we note the existence of a minimum value $K_{min}$ of the coupling strength $K$ for the anticipated synchronization to be stable (large values of the correlation coefficient). For a given $\theta$, the anticipation time reaches its highest value $\tau_{max}$ for a value of the coupling constant close to $K\gtrsim K_{min}$ and decreases monotonously with increasing $K$. This is so because for large $K$ values the feedback term induces a complex dynamics in the slave system reducing its possibility to synchronize with the master. It can also be seen that the stable region of synchronization (white area in Fig.\ref{f4}) is larger for $\theta> 0$, meaning that the use of an appropriate complex coupling parameter can enhance the stability of the synchronized solution in this system.

\begin{figure}[h]
\includegraphics[width=1.1\linewidth]{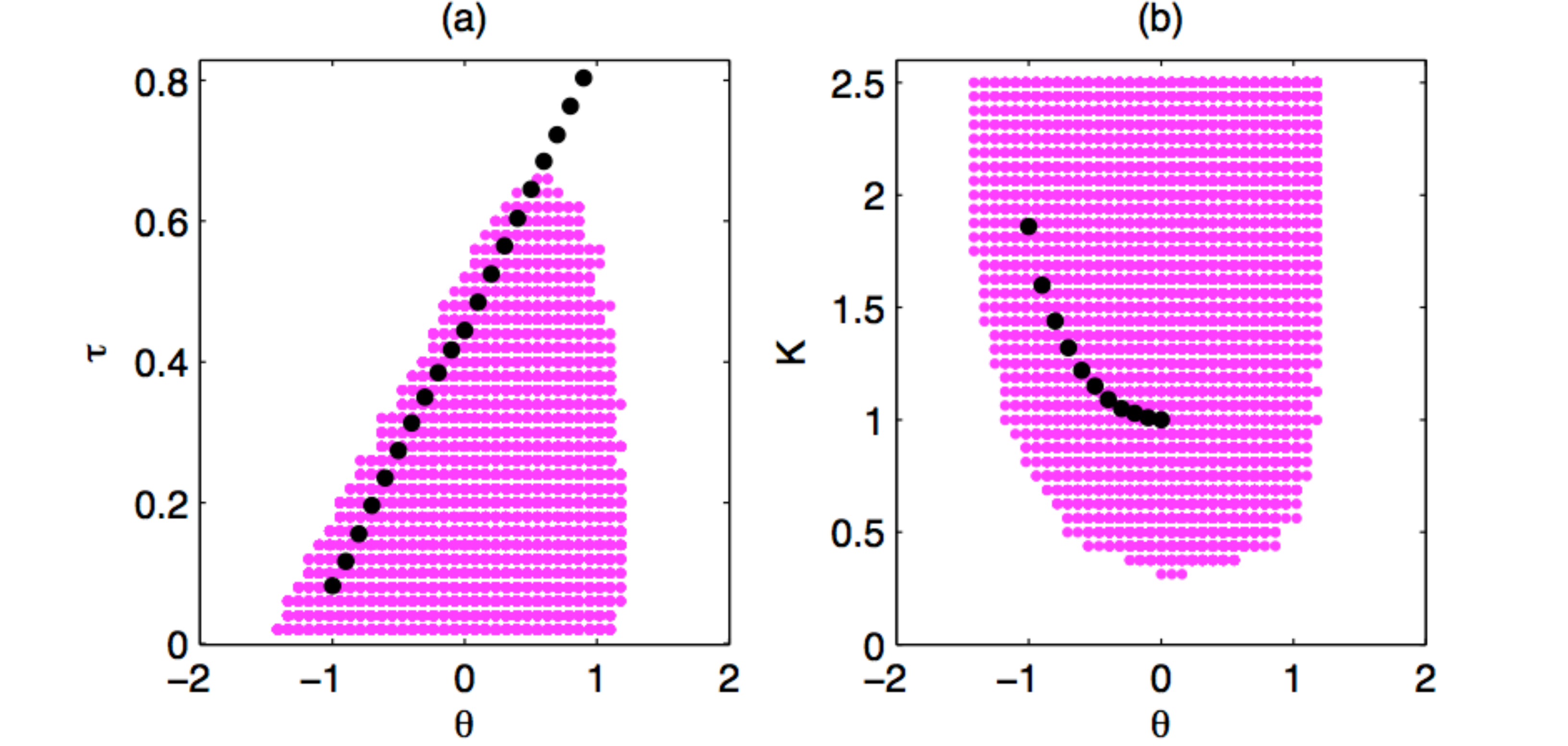}
\caption{\label{f5} (color online) Purple dots indicate the projections in the planes (a) $(\tau,\theta)$ and (b) $(K,\theta)$ of the region for which stable anticipated synchronization solutions have been found numerically in the defect turbulence regime with parameters $(c_1,c_2)=(3,-2.5)$. The black-dotted lines are the analytical results obtained from an approximate linear stability analysis, see Eq.(\ref{bifurcation}). }
\end{figure}

The stability regions in terms of the $\theta$ parameter, as a
function of $K$ and $\tau$, were also analyzed in detail. In
Fig.~\ref{f5} we plot projections of the region where stable
anticipated synchronization is obtained in the $(\theta,\tau)$ and
$(\theta,K)$ planes. From Fig.~\ref{f5}(a) we see that the largest
anticipation time $\tau_{max}\approx 0.6$ occurs for a non-zero
value of the phase coupling constant, $\theta \approx
\frac{\pi}{6}$. This corroborates the fact that complex coupling
increases the stability of the anticipated synchronization. It can
also be seen in this figure that the stability region is
asymmetric, tilted towards the positive values of $\theta$. This
asymmetry will be explained in the next section when we perform a
linear stability analysis of the anticipated solution. From
Fig.~\ref{f5}(b), it can be seen that the minimum coupling value
for which the anticipated synchronization is stable also depends
on $\theta$. For this particular set of parameters we find
$K_{min}\approx 0.3$ and occurs for $\theta\approx 0$.

In summary, our numerical analysis indicates that in the case of
defect turbulence regime the maximum anticipation time is
$\tau_{max}\approx 0.6$ for $\theta \approx \pi/6$ and $K=0.6$.
Keeping constant those values of $K$ and $\theta$, and moving to
the bichaos regime, $(c_1,c_2)=(1.1,-1.2)$, the maximum
anticipation time increases to $\tau _{max}\approx 1.6$, while in
the regime of phase turbulence, $(c_1,c_2)=(1.5,-0.8)$, we find
$\tau _{max}\approx1.9$, in agreement with the arguments exposed
before about the similarity between the anticipation times and the
linear correlation times.

The previous analysis has been restricted to a set of parameters in the defect turbulence regime. In bichaos and phase turbulence regimes the minimal coupling required for anticipated synchronization to occur is smaller. The minimum value of $K$ can be compared with the one necessary to observe phase synchronization in unidirectionally coupled CGL equations performed by Junge and coauthors \cite{junge}. In the latter, $K_{min}=0.1$ for the phase turbulence regime and increases for bichaos and defect turbulence regimes. Thus the tendency of the coupling value to grow when entering into more chaotic regimes is present in the synchronization scenario. 

\begin{figure}[h]
\includegraphics[width=1.1\linewidth]{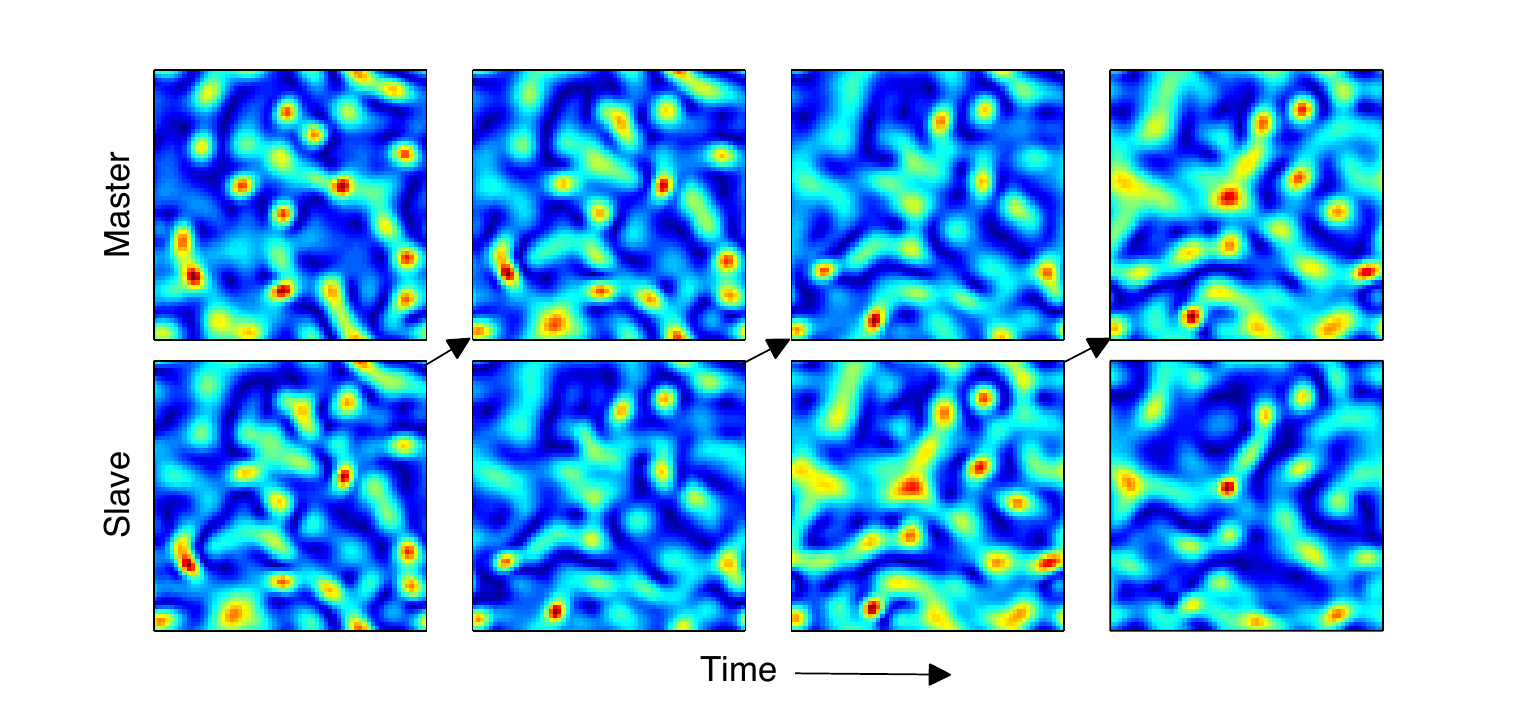}
\caption{\label{f6}(color online) Amplitude evolution of two-dimensional complex Ginzburg-Landau
equations, Eqs. (\ref{eq5dim1}-\ref{eq5dim2}), master (upper
row) and slave (lower row) for coupling parameters $\tau=0.37$,
$K=1.35$ and $\theta=-\frac{\pi}{4}$. Snapshots are shown at times
separated by a time unit $\tau$. Anticipated synchronization is
apparent since each frame in the dynamics of the master, upper
row, has been reached in an earlier frame of the slave system,
lower row, as indicated by the arrows. System size is $N\times
N=64\times 64$ with $\delta x=0.3$ and periodic boundary
conditions. Defect turbulence regime is considered with $(c_1,
c_2)=(3,-2.5)$.}
\end{figure}

As a final example, we study numerically the anticipated synchronization in the two-dimensional coupled CGL
equations described by:
\begin{eqnarray}
\dot{A}&=&\epsilon A+\alpha _1 \nabla^2 A -\alpha _2|A|^2A\label{eq5dim1}\\
\dot{B}&=&\epsilon B+\alpha _1 \nabla^2 B -\alpha _2|B|^2B+\kappa (A-B_\tau)\label{eq5dim2}
\end{eqnarray}
where $A=A(x,y,t)$, $B =B(x,y,t)$, $B_\tau=B(x,y,t-\tau)$, $\nabla^2 =\partial^2/\partial x^2 +\partial ^2/\partial y^2$. In Fig. \ref{f6} we show snapshots of the spatiotemporal evolution of the amplitude of the master (upper row) and slave (lower row) systems operating in the defect turbulence regime. Consecutive snapshots of the amplitudes are separated by a time $\tau$. It can be observed that the slave system anticipates the master by a time $\tau$ (indicated by diagonal arrows).

From our numerical results, we can infer that the
additional dimension makes the system more chaotic and thus, as
expected, the maximum anticipation time decreases as compared to the one-dimensional case. This is because the second spatial dimension increases the complexity of the system.

\section{Stability analysis}
We develop in this section a linear stability analysis of the anticipated synchronization solution $B(x,t)=A(x,t+\tau)$ of the one-dimension case, Eqs.(\ref{eq5}-\ref{eq6}). To this end we introduce
$\Delta (x,t) \equiv A(x,t)-B(x,t-\tau)$, which satisfies
\begin{equation}
\dot{\Delta}=\epsilon \Delta +\alpha_1 \Delta_{xx} - \kappa \Delta_\tau - \alpha_2 |A|^2 A +\alpha_2 |B_{\tau}|^2 B_\tau,
\end{equation}
with $\Delta_\tau=\Delta(x,t-\tau)$. Replacing $B_\tau=B(x,t-\tau)=A(x,t)-\Delta(x,t)$, and keeping only terms of first order in $\Delta$ we obtain:
\begin{equation}
\dot{\Delta}=\epsilon \Delta +\alpha_1 \Delta_{xx} - \kappa \Delta_\tau - 2 \alpha_2 |A|^2 \Delta - \alpha_2 A^2 \Delta^{\ast},
\end{equation}
Since $A^2$ is a highly oscillating term, we replace it by its average value $\langle A^2\rangle=0$. Hence, the last term of the previous equation vanishes, leading to a linear equation to determine the stability of the anticipated solution:
\begin{equation}
\dot{\Delta}=\epsilon \Delta +\alpha_1 \Delta_{xx} - \kappa \Delta_\tau - 2 \alpha_2 |A|^2 \Delta . \label{eq11}
\end{equation}
The aim is to determine if $\Delta(x,t)$ grows or decays to zero as a function of time.
To proceed, we make the further assumption that $A(x,t)$ is a plane wave of the form (\ref{eq3a}) and replace $|A|^2=\epsilon-q^2$. The resulting perturbation $\Delta^q(x,t)$ is expanded in Fourier modes, $\hat\Delta^q(l,t) \equiv \int \Delta^q(x, t) e^{-ilx} dx$, which satisfy
\begin{equation}
\dot{\hat{\Delta}^q}=[\epsilon - l^2 -2(\epsilon-q^2) -i (l^2 c_1 + 2 c_2 (\epsilon -q^2))] \hat\Delta^q- K e^{i\theta} \hat\Delta_{\tau}^q.\label{delta1}
\end{equation}
For this linear delay equation we make the typical ansatz \cite{erneux} that the solution is of the form $\hat\Delta ^q(l,t)=\exp(\lambda t)$
where $\lambda =\alpha +i\omega$. After replacing it in Eq.(\ref{delta1}) the real and imaginary parts lead to the following relations:
\begin{eqnarray}
\alpha &=&(\epsilon-l^2-2(\epsilon-q^2))-Ke^{-\alpha\tau}\cos(\omega\tau-\theta),\hspace{20pt}\\
\omega &=& -l^2 c_1-2 c_2(\epsilon-q^2)+K
e^{-\alpha\tau}\sin(\omega\tau-\theta).
\end{eqnarray}
The bifurcation points in the parameter space $(K,\theta,\tau)$ are obtained from the condition $\alpha=0$, i.e. when $\textrm{Re}[\lambda]$ changes from negative to positive. At the same time $\textrm{Im}[\lambda]$ is nonzero and the perturbation starts to oscillate and grow. Setting $\alpha=0$ we obtain for the stability of the $l$-th Fourier mode of the anticipated solution the following relation for $\tau$ as a function of
$K$ and $\theta$:
\begin{eqnarray}
\tau=\frac{\arccos (\frac{\epsilon - l^2 -2(\epsilon-q^2)}{K}) + \theta}{-l^2 c_1 -2 c_2 (\epsilon -q^2)\pm \sqrt{K^2-[\epsilon - l^2 -2(\epsilon-q^2)]^2}}.\hspace{10pt}
\label{bifurcation}
\end{eqnarray}
As the delay time that destabilizes the Fourier mode
$\hat\Delta^q(l,t)$ depends on $l$ and $q$, we search for the values
of $l$ and $q$ that yield the minimum value of $\tau$ (the maximum anticipation time). It turns
out that this minimum value occurs for $l=0$. In Fig.\ref{f4}
we plot the resulting stability lines in the $K-\tau$ plane.
We also plot in Fig.\ref{f5} the resulting curves for the maximum anticipation time $\tau$ for different values of
$\theta$ and compare them with the numerical values. A reasonable agreement only occurs for the left part of the stable region, failing to predict the decay of the stability for larger $\theta$ values. In the $K-\theta$ plane, we obtain a poor agreement for the analytical dependence of the stability region, indicating that the approximations are not appropriate in this case.

We now use this linear stability analysis to explain the lack of symmetry around $\theta=0$ observed in Fig.\ref{f5}(a). We consider a simple argument as the one used in the analysis of references \cite{pyragas2008,pyragas2010} for chaotic systems with time-delay coupling. From the plane-wave approximation used in the calculation, $A(x,t) \approx A_0(x)e^{i\omega t}$ we derive that $e^{i\theta}A(x,t)=A(x,t+\tau')$ with $\theta=\omega \tau'$, and a similar expression for $B(x,t)$. Consequently in Eq.(\ref{eq11}) the coupling term is transformed in the form $\kappa \Delta(x,t-\tau) \approx K \Delta(x,t-\tau'')$ with an effective delay time $\tau''=\tau-\tau'$. Therefore, for $\theta>0$, it is $\tau'>0$, reducing the effective delay time in the stability condition and hence increasing the stability of the synchronized solutions. On the contrary, if $\theta<0$, $\tau'<0$ and the effective delay increases reducing the stability of the synchronized solution.

\section{Conclusions}

In this paper we have presented some results on anticipated synchronization in two spatially extended complex Ginzburg-Landau systems, unidirectionally coupled in a master-slave configuration using a complex-valued coupling $\kappa=Ke^{i\theta}$ and a negative self-feedback delay term in the slave. Both in one and two spatial dimensions we have clearly observed the anticipation of the slave, which is able to reproduce the spatial patterns of the master an earlier time $\tau$, equal to the delay introduced in the self-coupling term of the slave.

Detailed results have been reported for the one-dimensional system in three different regimes of parameters, defect turbulence, bichaos and phase turbulence. We have found, in agreement with general arguments in systems with a small number of degrees of freedom, that the maximum anticipation time closely follows the linear autocorrelation time. The stability diagrams of this anticipated synchronization, defined as regions of high correlation between the master and the slave a time $\tau$ earlier, have been obtained in the parameter space $(K,\theta,\tau)$. We have observed that the value of $\theta$ is relevant to determine the stability region and that the stability curve in the parameter space $(\theta,\tau)$ is asymmetric being possible to reach larger anticipation times for positive values of $\theta$. Then, the consideration of a complex-valued coupling constant in the system appears to be (for some specific values) a way to increase the region of stability of the anticipated synchronization and reach larger anticipation times.

We have performed a linear stability analysis of the anticipated synchronization state and compared it with the numerical results in the one-dimensional system. We have obtained a qualitative good agreement for the prediction of the maximum anticipation time in the $(\tau,K)$ plane, but have failed in the estimation of the maximum coupling strength, indicating that the linear approximation is not valid in this case. 

Finally, it is worth mentioning that anticipated synchronization in spatially extended systems may be used in the prediction of e.g. the dynamics of chemical reactions. It could be interesting to compare the local (point-to-point) and global (all-to-all) types of coupling, since the latter might be more appropriate in real applications.

\section{Acknowledgments}

We acknowledge financial support from MINECO (Spain) and FEDER (EC) under project FIS2012-30634, and Comunitat Aut\`onoma de les Illes Balears. M.C. acknowledges Regione Toscana for financial support.

\end{document}